\documentclass[journal]{IEEEtran}
\usepackage{tabularx}  
\ifCLASSINFOpdf
\else
   \usepackage[dvips]{graphicx}
\fi
\usepackage{url}

\usepackage{amsmath,amssymb,graphicx,cite,url}
\usepackage{hyperref}
\usepackage{graphicx}
\usepackage{multirow}
\usepackage{booktabs}
\usepackage{threeparttable}
\usepackage{enumitem}
\usepackage{float}  %
\begin{document}

\title{A Dual-Stage Time-Context Network for Speech-Based Alzheimer’s Disease Detection}

\author{Yifan Gao, Long Guo and Hong Liu
\thanks{This work is supported by the Fundamental Research Funds for Central Universities under grants SCU2024D055 and SCU2024D059 }
\thanks{The authors are with the College of Computer Science, National Key Laboratory of Fundamental Science on Synthetic Vision, Sichuan University, Chengdu, 610065 China, (e-mail: 2023223045222@scu.stu.edu.cn; 2022223040087@stu.scu.edu.cn; liuhong@scu.edu.cn)}

}
\markboth{}
{Shell \MakeLowercase{\textit{et al.}}: Bare Demo of IEEEtran.cls for IEEE Journals}
\maketitle

\begin{abstract}
Alzheimer’s disease (AD) is a progressive neurodegenerative disorder that leads to irreversible cognitive decline in memory and communication. 
Early detection of AD through speech analysis is crucial for delaying disease progression. 
However, existing methods mainly use pre-trained acoustic models for feature extraction but have limited ability to model both local and global patterns in long-duration speech. 
In this letter, we introduce a Dual-Stage Time-Context Network (DSTC-Net) for speech-based AD detection, integrating local acoustic features with global conversational context in long-duration recordings.
We first partition each long-duration recording into fixed-length segments to reduce computational overhead and preserve local temporal details.
Next, we feed these segments into an Intra-Segment Temporal Attention (ISTA) module, where a bidirectional Long Short-Term Memory (BiLSTM) network with frame-level attention extracts enhanced local features.
Subsequently, a Cross-Segment Context Attention (CSCA) module applies convolution-based context modeling and adaptive attention to unify global patterns across all segments.
Extensive experiments on the ADReSSo dataset show that our DSTC-Net outperforms state-of-the-art models, reaching 83.10\% accuracy and 83.15\% F1.

\end{abstract}


\IEEEpeerreviewmaketitle

\section{Introduction}
\label{sec:intro}

\IEEEPARstart{A}{lzheimer’s} disease (AD) is a progressive neurodegenerative disorder, with a rapidly increasing prevalence around the world as populations age.
Over 55 million people worldwide are affected by AD \cite{5000W_AD}, creating a pressing healthcare burden.
Patients typically show significant symptoms such as memory decline, speech impairments, and impaired executive functioning, severely compromising their quality of life.\cite{Symptons}%
Although no definitive cure exists, early detection and intervention can substantially slow disease progression \cite{EarlyDiagnosis1,EarlyDiagnosis2}.
Therefore, developing cost-effective and scalable AD screening methods is a urgent goal.

Compared with traditional diagnostic methods such as brain imaging or blood tests \cite{Detect_Method}, speech-based approaches are more affordable, non-invasive, and suitable for remote use. These advantages have driven rapid growth in this research area, particularly because early AD symptoms often involve subtle shifts in speech rate, pauses, word retrieval, and expression coherence \cite{speakfeature-AD1}, \cite{speakfeature-AD2}, \cite{speakfeature-AD3}. 
Such changes are captured by acoustic features like prosody, fluency, and pause patterns, enabling earlier and more robust detection. Prior studies confirm that these features can effectively distinguish AD patients from healthy controls and have strong discriminative power in early-stage diagnosis \cite{egemaps-luwen}, \cite{traditional1}, \cite{traditional2}, \cite{traditional3}.

With recent advances in deep learning, pre-trained acoustic models like Wav2Vec2.0 \cite{Wav2Vec}, HuBERT \cite{Hubert}, and Whisper \cite{openai-Whisper} have shown remarkable effectiveness in extracting high-level speech representations.
To learn broadly applicable acoustic representations, these models often leverage advanced architectures (e.g., Transformers or convolution-Transformer hybrids) and rely on self-supervised or weakly supervised training over large-scale audio corpora.
Empirical evidence indicates that integrating the hidden representations learned by pre-trained models with traditional features \cite{luz24}, or further fine-tuning models \cite{whipser-transfer}, often achieves substantial improvements in tasks like speech classification, emotion recognition, and semantic comprehension \cite{w2v-cls}.
Specifically, for AD detection, the multi-level and high-quality feature embeddings from pre-trained models have shown promise for enhancing both accuracy and robustness \cite{other-Hubert-w2v-embedding1}, \cite{other-Hubert-w2v-embedding2}, \cite{other-Hubert-w2v-embedding3}.

Despite the computational challenges of using pre-trained acoustic models on long-duration AD speech, these models still appeared in datasets such as ADReSS \cite{adress}, ADReSSo \cite{adresso}, and DementiaBank \cite{DB}. In these datasets, the average speech length often exceeded 60 seconds, making it hard to capture both local acoustic details and global dialogue patterns.

Wav2Vec2.0 and HuBERT did not strictly limit input length but had to store intermediate states for the entire sequence, which could greatly increase memory usage. Thus, researchers often split recordings into short segments. Whisper used convolution-based downsampling to reduce computation but was pre-trained on 30-second inputs and also segmented longer speech by default.

Previous studies approached segment-based processing in different ways. For example, Gauder et al. \cite{segment-w2v}, Ying et al. \cite{w2v-segement-average1}, and Pérez-Toro et al. \cite{2024-top1} divided long audio into short segments to reduce computational load, but this partitioning disrupted acoustic and semantic continuity across segment boundaries. Balagopalan et al. \cite{w2v-segement-average2} and Favaro et al. \cite{JHU} then averaged features across segments, overlooking variations that could reveal changes in speech rate or pauses. Li et al. \cite{cunk-whipser-encoder} introduced an attention mechanism at the frame level but still performed segment averaging in the end, thereby losing cross-segment context. Meanwhile, other work \cite{asr1}, \cite{asr2}, \cite{asr3} relied heavily on Whisper for automatic speech recognition, focusing primarily on textual transcriptions and overlooking important acoustic markers of AD.

These approaches reduce computational overhead but often overlook both local acoustic details and broader conversational patterns. Accordingly, more refined segmentation and better context modeling are needed to preserve subtle acoustic nuances and global information crucial for AD detection.

Building on previous work, this study partitions long-duration speech into fixed-length segments to better capture both local acoustic features and global conversational patterns. 
Further, we propose a novel architecture for AD detection. Our major contributions are listed below:

\begin{enumerate}[leftmargin=*]
  \item We first segment the long-duration audio at the outset and extract multi-level embeddings from each segment, capturing more detailed temporal and acoustic features that are crucial for early AD detection.
  \item We propose a novel architecture, named Dual-Stage Time-Context Network (DSTC-Net), consisting of two key modules: the Intra-Segment Temporal Attention (ISTA) module, which integrates a BiLSTM with frame-level attention to refine local features, and the Cross-Segment Context Attention (CSCA) module, which employs convolution-based context modeling and adaptive attention distribution to fuse global conversational patterns.
  \item Extensive experiments on the ADReSSo dataset demonstrate that DSTC-Net outperforms recent state-of-the-art (SOTA) methods, achieving 83.10\% accuracy and 83.15\% F1, effectively balancing local acoustic features with broader discourse context for reliable AD screening.
\end{enumerate}

\begin{figure*}[ht]
    \centering
    \includegraphics[width=\textwidth]{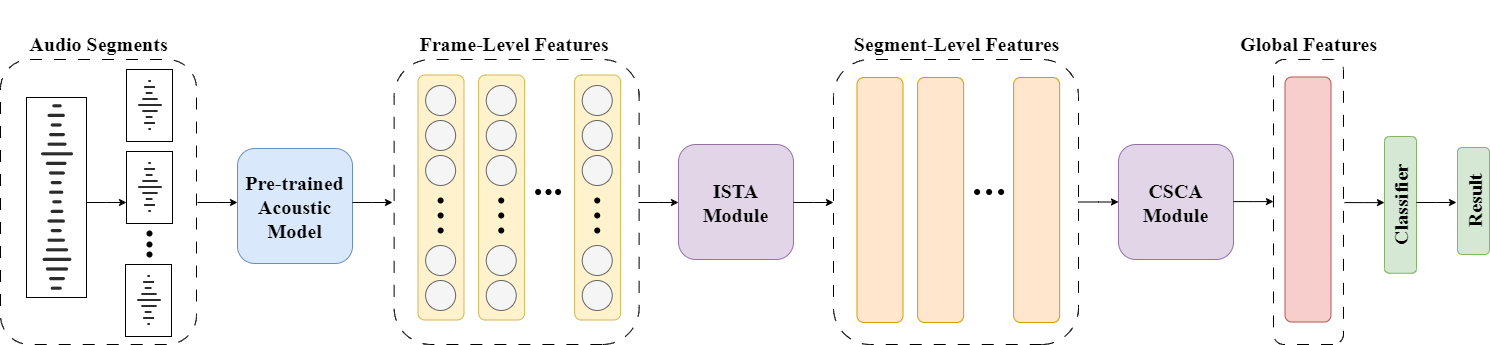}
    \caption{ Overview of the proposed DSTC-Net framework.}
    \label{fig:Whisper}
\end{figure*}

\section{Method}

\subsection{Audio Preprocessing}

\subsubsection{Noise Reduction}
In real-world conversations, environmental and microphone noise can weaken speech signals and lower analysis accuracy. To solve this problem, we use the Koala\footnote{Koala: \href{https://picovoice.ai/docs/api/koala-web/}{https://picovoice.ai/docs/api/koala-web/}} tool to reduce noise. 
It strengthens the main speech frequencies and makes feature extraction clearer. Previous studies \cite{2025-top-1} show that moderate noise reduction not only makes features easier to separate but also improves classification performance.

\subsubsection{Segmentation}
In this study, we partition the audio data into fixed-length segments with about 10\% overlap between adjacent segments. 
On the one hand, this segmentation facilitates evaluating how different segment lengths impact model performance; on the other hand, it preserves contextual and prosodic features near segment boundaries, preventing abrupt interruptions from hard cuts. 
Additionally, segments shorter than the fixed length are zero-padded to ensure uniform input dimensions.

\begin{table}[!t]
\centering
\caption{Configuration of Pre-trained Acoustic Models}
\label{table:speech-models}
\small
\setlength{\tabcolsep}{5pt} 
\begin{tabular}{c c c c c} 
\toprule
\textbf{Model} & \textbf{Size} & \textbf{Parameters} & \textbf{Layers} & \textbf{Width} \\
\midrule
Hubert      & base  & 94.7\,M   & 12  & 768 \\
Wav2Vec~2.0 & base  & 95.0\,M   & 12  & 768 \\
Whisper     & small & 244.2\,M  & 12  & 768 \\
\bottomrule
\end{tabular}
\end{table}

\subsection{Pre-trained Acoustic Model}
We compare three pre-trained speech models for acoustic feature extraction: Wav2Vec 2.0, Hubert, and Whisper.

Both Wav2Vec 2.0 and Hubert leverage self-supervised learning (SSL) to derive high-level, context-rich representations from large volumes of unlabeled speech data. Specifically, Wav2Vec 2.0 employs multi-layer convolutional neural networks (CNNs) to encode raw audio into latent representations, and then uses contrastive learning to predict masked time steps, effectively modeling rich contextual dependencies . In contrast, Hubert introduces a distinct approach by using an offline clustering strategy. This strategy generates pseudo-labels for unlabeled audio through clustering, guiding the model to predict cluster assignments for masked segments, thereby enabling the model to learn more stable and robust acoustic-language features .

In contrast, Whisper is trained in a weakly supervised, multi-language, multi-task framework, using a sequence-to-sequence approach on 680,000 hours of speech–text alignment data. This extensive training leads to stronger robustness and generalization in multilingual or noisy environments.

In our experiments, we use the base versions of Wav2Vec 2.0 and Hubert, and the small version of Whisper, as summarized in Table \ref{table:speech-models}. 
All three models share the same number of encoder layers, enabling direct comparisons across layers.

Although structurally and output-wise distinct, each model converts input audio into a four-dimensional feature tensor $(L_a, N_a, T_a, F_a)$. This tensor represents the number of encoder layers, segments, time steps, and feature dimensions, respectively, thereby providing unified acoustic embeddings for downstream tasks such as AD detection.

\begin{figure}[htbp]
    \centering
    \begin{minipage}[b]{0.48\linewidth}
        \centering
        \includegraphics[height=0.2\textheight]{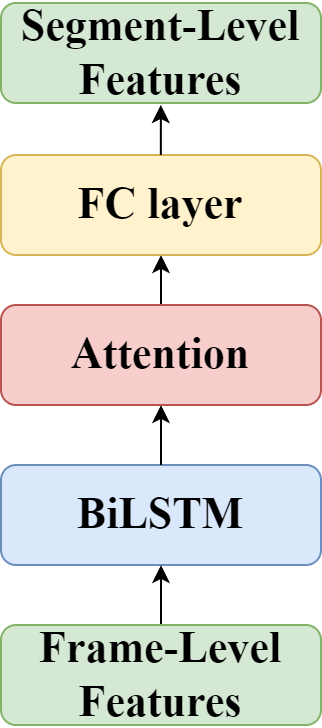}
        \caption{ISTA Module.}
        \label{fig:sam}
    \end{minipage}
    \hfill
    \begin{minipage}[b]{0.48\linewidth}
        \centering
        \includegraphics[height=0.2\textheight]{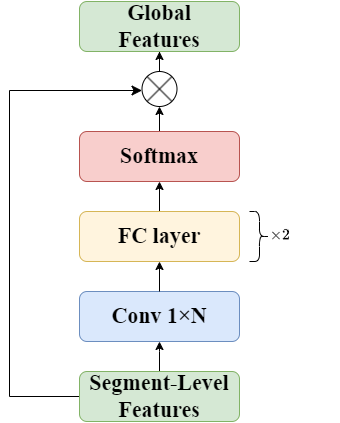}
        \caption{CSCA Module.}
        \label{fig:gam}
    \end{minipage}
\end{figure}

\subsection{Dual-Stage Time-Context Network}
As shown in Fig.~\ref{fig:Whisper}, the proposed Dual-Stage Time-Context Network (DSTC-Net) captures both intra-segment dependencies and cross-segment correlations for AD detection. It comprises two main modules: (1) an Intra-Segment Temporal Attention (ISTA) module, which uses a BiLSTM network and attention to highlight critical frames within each segment, and (2) a Cross-Segment Context Attention (CSCA) module, which integrates segment-level features into a global representation for final classification.

\subsubsection{Intra-Segment Temporal Attention Module}
\label{sec:ista}

As illustrated in Fig.~\ref{fig:sam}, the ISTA module first extracts frame-level features from pre-trained acoustic models (e.g., Wav2Vec~2.0, Hubert, or Whisper). Although these features provide rich speech information, the temporal dynamics of disease-related cues are not sufficiently captured. Therefore, ISTA feeds these features into a BiLSTM to model forward and backward dependencies, generating a hidden state at each time step.

Given an acoustic-based feature or embedding vector 
$X = [x_1, x_2, \dots, x_N] \in \mathbb{R}^{T \times F}$
as the input sequence, where $T$ is the number of time steps (varying with audio length) and $F$ is the feature dimension, we obtain the hidden state $\displaystyle h_t \in \mathbb{R}^{256}$ at time step $t$ by concatenating the forward hidden state $\displaystyle h_{t}^{f} \in \mathbb{R}^{128}$ and the backward hidden state $\displaystyle h_{t}^{b} \in \mathbb{R}^{128}$:
\begin{equation}
\label{eq:bilstm}
h_t = h_{t}^{f} \oplus h_{t}^{b}, {\quad t = 1,2,\dots,T}.
\end{equation}

However, mere sequence modeling may overlook crucial frames for AD detection.
Therefore, we adopt an attention mechanism~\cite{Attention} after the BiLSTM to assign varying weights to each time step. Specifically, we compute the similarity between the final BiLSTM hidden state (acting as a segment-level “summary”) and each time step’s hidden state:
\begin{equation}
\label{eq:attention}
\alpha_t = \mathrm{softmax}\bigl(h_{T}^{\top} \, W \, h_t\bigr),
\end{equation}
where $h_{T}$ is the hidden state at the final time step of the BiLSTM, $h_t$ is the hidden state at time $t$, and $W$ is a learnable parameter. A larger attention weight $\alpha_t$ implies that the corresponding time step contributes more to AD discrimination.

Next, we form a weighted sum of all time-step states, capturing both the global context and critical frames. Finally, this attention-derived vector is concatenated with the BiLSTM outputs and passed through a fully connected layer (e.g., with a $\tanh$ activation). The resulting segment-level representation thus retains overarching temporal dependencies while focusing on key-frame information essential for AD detection.

\subsubsection{Cross-Segment Context Attention Module}

Building upon the segment-level representations from ISTA, the CSCA module shown in Fig.~\ref{fig:gam} coordinates global context aggregation by applying segment-level attention across all segments.

It then employs fully connected layers for nonlinear mapping, followed by a softmax function to derive each segment’s relative weight in the global classification task.

Finally, we aggregate these segment vectors via a weighted sum according to the attention weights, thereby forming the final global feature vector $G$:
\begin{equation}
G = \sum_{m=1}^{M} \beta_m \, z'_m
\end{equation}
where $\beta_m$ is the attention weight for the $m$-th segment, and $z'_m$ is the segment-level representation after convolution. 
This mechanism highlights discriminative segments and fuses inter-segment correlations, offering a more comprehensive global representation for AD recognition.

\subsubsection{Classifier}
After deriving the global feature vector $G$, we apply a simple fully connected layer to determine whether each speech sample originates from an individual with AD.

\begin{figure*}[ht]
\centering
\begin{minipage}{0.32\textwidth}
\centering
\includegraphics[width=\linewidth]{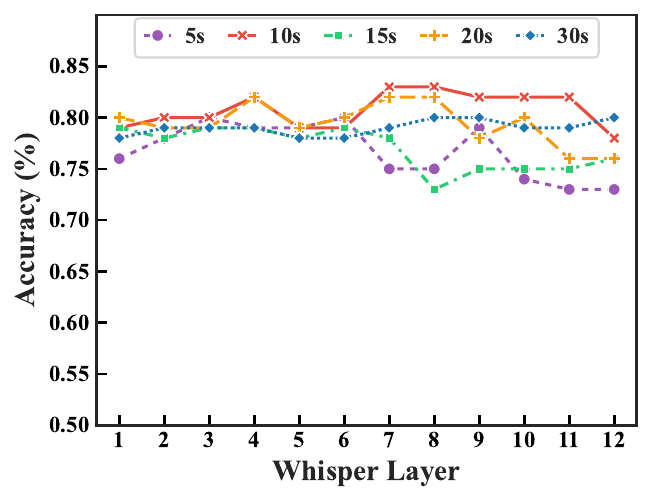}
\end{minipage}
\hfill
\begin{minipage}{0.32\textwidth}
\centering
\includegraphics[width=\linewidth]{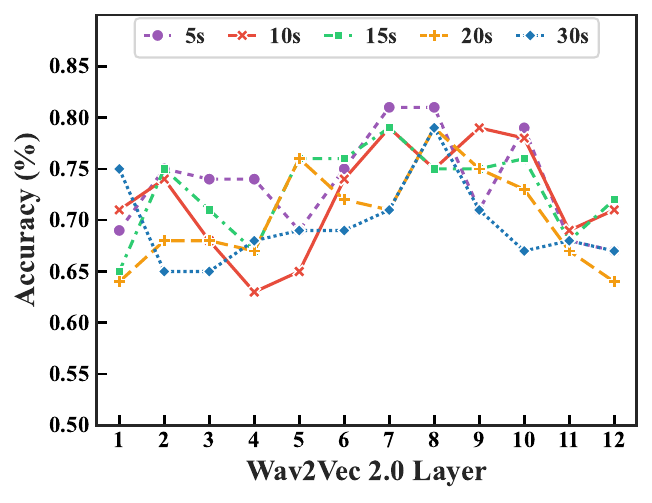}
\end{minipage}
\hfill
\begin{minipage}{0.32\textwidth}
\centering
\includegraphics[width=\linewidth]{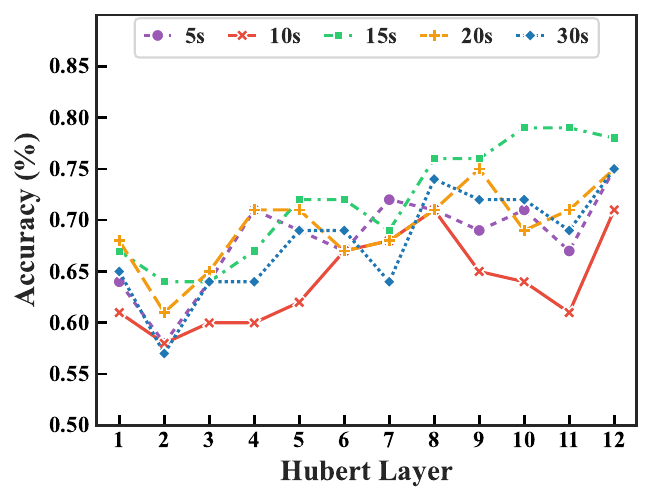}
\end{minipage}
\caption{The relationship between the contextual layer depth and the performance of the extracted features for different segmentation lengths; results are obtained from Whisper (left), Wav2Vec2.0 (middle), and Hubert (right) on the ADReSSo dataset.}
\label{fig:result}
\end{figure*}

\section{Experiments}
\label{sec:experiments}
\subsection{Datasets}
This study employs the standardized corpus from the 2021 Alzheimer’s Dementia Recognition through Spontaneous Speech (ADReSSo) challenge, derived from the “Cookie Theft” picture description task in the Pitt DementiaBank database. 
The dataset comprises speech recordings and transcripts from 237 English-speaking participants (122 with AD, 115 healthy controls). 
On average, AD samples last 65.70~$\pm$~38.3 seconds, while HC samples last 61.60~$\pm$~26.9 seconds. 

\subsection{Implementation Details}
Our model was implemented in PyTorch~2.0.1 on an L20 GPU. The ISTA module uses a BiLSTM (128 hidden units per direction, concatenated to 256) with a 0.3 dropout, followed by a multiplicative attention and a 128-dimensional fully connected layer. The CSCA module has two fully connected layers (64 and 32), culminating in a linear classifier. We trained using Adam (learning rate = 1e-4, batch size = 32) for up to 200 epochs, applying early stopping if validation did not improve after 50 epochs. The dataset adhered to its predefined partition, with model selection via 10-fold cross-validation and final evaluation on the held-out set.

\section{Results}
Fig \ref{fig:result} illustrates how different segmentation lengths and encoder depths jointly affect the recognition accuracy of Whisper, Wav2Vec 2.0, and Hubert.

In the left sub-figure (Whisper), accuracy remains high across all segment lengths, with only minor fluctuations. The best performance appears around layers 6–10 when using 10 s segments, suggesting that 10 s is the most effective duration for Whisper. Even the shortest (5 s) or longest (20 s/30 s) segments deliver strong results, highlighting Whisper’s robust feature representation.

In the middle sub-figure (Wav2Vec 2.0), accuracy dips slightly at layers 1–2 and 11–12 but peaks in the mid-layers (especially layers 7–8). Of the five tested durations, 5 s yields the highest accuracy in these mid-layers, indicating that shorter segments best leverage Wav2Vec 2.0’s architecture.

In the right sub-figure (Hubert), accuracy steadily improves with deeper layers. Early layers (1–3) remain moderate, but the final layers (10–12) show notable gains. At these depths, 15 s segments achieve the strongest performance, suggesting Hubert benefits from a longer context window.

Overall, each model favors a different segmentation length: 10 s for Whisper, 5 s for Wav2Vec 2.0, and 15 s for Hubert. These preferences likely arise from differences in architecture, pretraining strategies, and context requirements.

\begin{table}[ht]
\centering
\caption{Comparison of Different Acoustic Feature Extraction Methods}
\label{tab:comparison}
\small
\setlength{\tabcolsep}{3pt} 
\begin{tabular}{>{\centering}p{0.18\textwidth} >{\centering}p{0.13\textwidth} cc}
\toprule
\textbf{Method} & \textbf{Model} & \textbf{Accuracy} & \textbf{F1} \\
\midrule
Luz et al.\cite{adresso} & eGeMAPS & 64.79 & -- \\
Balagopalan et al.\cite{w2v-segement-average2} & Wav2Vec 2.0 & 67.61 & 70.89 \\
Pappagari et al.\cite{Pappagari} & x-vectors & 74.70 & 74.50 \\
Li et al.\cite{Whisper-fullASR} & Whisper & 77.46 & 77.46 \\
Pan et al.\cite{pan} & Wav2Vec 2.0 & 78.87 & 78.49 \\
Gauder et al.\cite{segment-w2v} & Wav2Vec 2.0 & 78.87 & -- \\
Wang et al.\cite{wang} & Hubert & 79.05 & 81.07 \\
\midrule 
\multirow{3}{*}{\textbf{Proposed}} 
& Hubert& 79.12 & 78.95 \\
& Wav2Vec2.0& 81.69 & 80.12 \\
& \textbf{Whisper} & \textbf{83.10} & \textbf{83.15} \\
\bottomrule
\end{tabular}
\end{table}

Table \ref{tab:comparison} shows that among our three acoustic-only approaches using Wav2Vec2.0, Hubert and Whisper, the Whisper-based version achieves the highest accuracy of 83.10\% and an F1-score of 83.15\% on the ADReSSo dataset. 
These results highlight the effectiveness of leveraging robust pre-trained acoustic representations and carefully designed segmentation strategies. 
The architectural refinements and multi-scale contextual fusion in our DSTC-Net enable it to capture both local and global speech features, driving state-of-the-art performance for AD detection.

\begin{table}[ht]
\centering
\caption{COMPARISON OF THE OVERALL ACCURACY AND F1 OF ABLATED MODELS}
\label{tab:ablation}
\setlength{\tabcolsep}{8pt} 
\renewcommand{\arraystretch}{1.2} 
\begin{tabularx}{0.45\textwidth}{c c c >{\centering\arraybackslash}X >{\centering\arraybackslash}X} 
\toprule
\textbf{Model} & \textbf{ISTA} & \textbf{CSCA} & \textbf{Accuracy} & \textbf{F1} \\
\midrule
\multirow{4}{*}{\textbf{Whisper}} 
            &               &               & 77.46            & 77.56      \\
            & \checkmark    &               & 81.69            & 80.79      \\
            &               & \checkmark    & 80.12            & 79.61      \\
            & \checkmark    & \checkmark    & \textbf{83.10}            & \textbf{83.15}      \\
\bottomrule
\end{tabularx}
\end{table}

Table \ref{tab:ablation} summarizes the Whisper model’s performance with various module configurations.
Both the Intra-Segment Temporal Attention (ISTA) and Cross-Segment Context Attention (CSCA) modules individually enhance performance over the baseline, while their combination achieves the highest accuracy (83.10\%) and F1-score (83.15\%). 
These results confirm that integrating both local feature refinement and global context fusion is essential for optimal performance.

\section{Conclusion}
\label{sec:conclusion}
In this letter, we introduced a Dual-Stage Time-Context Network (DSTC-Net) specifically designed for AD detection from long-duration speech. 
By segmenting speech to capture local acoustic features and integrating an Intra-Segment Temporal Attention (ISTA) module, we enhance fine-grained pattern modeling within segments. 
Simultaneously, the Cross-Segment Context Attention (CSCA) module aggregates broader conversational features, effectively capturing global discourse patterns crucial for robust AD screening.
Experiments on the ADReSSo dataset demonstrate that DSTC-Net achieves 83.10\% accuracy and an 83.15\% F1 score, surpassing recent state-of-the-art approaches. 
These findings confirm the value of modeling both local and global speech patterns in long-duration data, illustrating DSTC-Net’s promise for early and reliable AD detection.

\clearpage              
\bibliographystyle{IEEEtran}  
\bibliography{main}           
\end{document}